\documentclass[12pt,letterpaper]{JHEP3}         
\usepackage{epsfig}

\def\be{\begin{eqnarray}}
\def\ee{\end{eqnarray}}
\newcommand{\nn}{\nonumber}
\newcommand\para{\paragraph{}}

\newcommand{\eqn}[1]{(\ref{#1})}

\def\Dslash{\,\,{\raise.15ex\hbox{/}\mkern-12mu D}}
\def\Dbarslash{\,\,{\raise.15ex\hbox{/}\mkern-12mu {\bar D}}}
\def\delslash{\,\,{\raise.15ex\hbox{/}\mkern-9mu \partial}}
\def\delbarslash{\,\,{\raise.15ex\hbox{/}\mkern-9mu {\bar\partial}}}
\def\pslash{\,\,{\raise.15ex\hbox{/}\mkern-9mu p}}
\def\calDslash{\,\,{\raise.15ex\hbox{/}\mkern-12mu {\cal D}}}

\newcommand{\Tr}{{\rm Tr}}

\def\lae{\mathrel{\mathop{\smash{\lower .5 ex \hbox{$\stackrel<\sim$}}}}}
\def\lae{\mathrel{\mathop{\smash{\lower .5 ex \hbox{$\stackrel>\sim$}}}}}

\title{Universal Resistivity from Holographic Massive Gravity}

\author{Mike Blake and David Tong\\

 Department of Applied Mathematics and Theoretical Physics\\
 University of Cambridge, UK
\\ {\ } \\
{\tt m.a.blake, d.tong@damtp.cam.ac.uk}
}

\abstract{Massive gravity provides a holographic model for theories exhibiting momentum dissipation. We provide an analytic expression for the DC conductivity. The result is universal, depending only on properties of the infra-red horizon, and holds at finite temperature and  charge density.   In addition, we provide a derivation of black hole thermodynamics in holographic massive gravity and show that the resulting physics is sensible.}

\begin{document}
\pdfoutput=1
\pagestyle{plain} \setcounter{page}{1}
\newcounter{bean}
\baselineskip16pt

\section{Introduction}

In the presence of a non-vanishing charge density, any system with a conserved momentum current will exhibit a divergent conductivity at low frequencies. The physics behind this is straightforward: a constant electric field causes the charges to accelerate but, with no mechanism to lose momentum, there can be no current dissipation. 

\para
In materials, the presence of impurities or a background lattice structure means that momentum is not conserved (or is conserved only modulo reciprocal lattice vectors). This ensures that the DC conductivity remains finite. However any attempt to model systems using translationally invariant quantum field theories will run into problems unless the effects of momentum dissipation can be incorporated. 

\para
In the framework of holography, there have been a number of methods employed to include momentum dissipation and extract the DC conductivity. Conceptually, they fall into two classes. In the first method, one considers a parametrically small number of charged degrees of freedom, immersed in a bath of neutral degrees of freedom which can absorb the momentum. This  approach underlies the probe brane \cite{aob,hpst} and probe fermion \cite{fermi1,fermi2} calculations of conductivity. 

\para
The second method is to  implement the effect of translational symmetry breaking in the holographic context. This can be done 
perturbatively, viewing the lattice as an irrelevant operator from the IR perspective \cite{sandiego}, or 
more directly by computing charge transport in the background of a spatially modulated bulk  \cite{jorge1,jorge2,jorge3,aristos} or in the presence of impurities \cite{impure1,pauli}.



\para 
Recently, an alternative approach to holographic momentum dissipation was suggested by Vegh \cite{vegh}. This approach doesn't make use of any specific mechanism, but instead aims to provide an effective bulk description of a theory that does not conserve momentum. The basic idea is simple: 
conservation of the stress-energy tensor in the boundary theory arises due to diffeomorphism invariance in the bulk. If we want to model a theory without momentum conservation, we must work with a gravitational theory without diffeomorphisms. Modifications of general relativity which break diffeomorphism invariance go by the name of {\it massive gravity}. 

\para
Holographic massive gravity does not describe a Hamiltonian boundary theory (because, for example, it has translational invariance but no momentum conservation). The hope is that massive gravity can be thought of as a coarse-grained, low-energy bulk description of some (perhaps any?)  system with momentum dissipation, such as that induced by impurities or a background lattice. Clearly a pressing issue is to understand how, if at all,  holographic massive gravity can be derived from general relativity in AdS. There has been earlier work along these lines in \cite{elias,andreas,poratti} in which two AdS bulks are coupled through double-trace operators, generating a mass for a linear combination of the gravitons. However, it is currently unclear if this approach -- which is reminiscent of the  coupling to  a ``neutral bath" sector described in above -- gives rise to the same class of ghost-free massive gravity theories proposed in \cite{vegh}.

\para
Our purpose in this paper is to study the conductivity in massive gravity. Indeed, it was shown in \cite{vegh} that the DC conductivity is finite, while the low-frequency behaviour exhibits a familiar Drude peak. Further aspects of both conductivity and hydrodynamics have been explored by Davison \cite{davison}. In particular, at suitably high temperatures the continuity equation for the momentum current becomes, 
\be \partial_i T^{ij} = -\tau^{-1} T^{tj}\label{momdis}\ee
where $j$ is a spatial index as befits a momentum current. The scattering time $\tau$ is related to the graviton mass. 

\para
Here, we show that the DC conductivity can be calculated analytically as a function of the temperature and charge density. The result is universal: it depends only on properties of the black hole horizon, not on the bulk geometry. Specifically, the DC conductivity depends on the horizon radius and the value of the graviton mass at the horizon.   From the DC conductivity, we can extract a scattering time $\tau$: it agrees with the hydrodynamic scattering time that arises in \eqn{momdis}.

\para 
Our method follows the membrane paradigm approach of \cite{il}. We will show that, despite the graviton gaining a mass,  there remains a ``massless mode" in the bulk, a linear combination of the graviton and gauge field.  This massless mode encodes the information about the conductivity and, in the zero frequency limit, does not evolve in the radial direction.

\para
The paper is organised as follows. In Section 2, we review the ghost-free holographic massive gravity proposal of \cite{vegh}. The meat of the paper is contained in Section 3. After a detailed review of the results of \cite{il} applied to conductivity, we turn to the more complicated case of conductivity in massive gravity. Equation \eqn{result} for the DC conductivity is the main result of the paper. In Section 4 we study theories with a dilaton coupling and show that the result for the DC conductivity remains essentially unchanged, although the different scaling of the horizon radius with temperature can lead to qualitatively different behaviours for the conductivity. Section 5 contains an extended discussion in which we offer some speculations on the uses of massive gravity. We also include an appendix in which we compute the thermodynamics of black holes in massive gravity; we show that familiar results such as the Bekenstein-Hawking entropy and the first law of thermodynamics continue to hold.


\section{Massive Gravity}

In this section we introduce the basic features of massive gravity and the associated background solutions. 
Our starting point is the familiar Einstein-Hilbert-Maxwell action in $d=3+1$ dimensions with a negative cosmological constant,
\be S_1 = \int d^4x\sqrt{-g}\  \left[\frac{1}{2\kappa^2}\left(R
+\frac{6}{L^2}\right)-\frac{1}{4e^2}F_{\mu\nu}F^{\mu\nu} \right]\label{action1}\ee
In what follows, we will work with the
usual coordinates $(t,r,x,y)$ that label the Poincar\'e patch of AdS${_4}$, with the boundary at $r=0$. 
(Although, as we'll see, AdS is no longer a solution once the graviton has a
mass).

\para
Augmenting the Einstein-Hilbert action with a mass term for the graviton is not a simple affair. Even
at the linearised level, a Fierz-Pauli mass term requires a fine-tuning to ensure that no ghost field --- a scalar with wrong sign kinetic term ---
propagates. However, ghosts are not so easily exorcised. At the non-linear  level, one typically finds that the ghost reappears when looking at excitations around non-trivial backgrounds \cite{bd}. A nice review of these issues can be found in
\cite{hinter}.

\para
Recently, a class of non-linear massive gravity theories has been proposed \cite{derahm1,derahm2} in which the ghost field is absent \cite{rosen1,rosen2}.  This is the theory we work with here.

\para
(We note that the theory is not immune from difficulties. Even before worrying about quantum issues, at the classical level it is known that the theory suffers from superluminal propagation \cite{deser} while it has been shown that, around a branch of cosmological solutions, the ghost may reappear, now sitting within the five massive spin 2 degrees of freedom \cite{niz,shinji}. Neither of these issues seem particularly pressing for a holographic theory describing, say, a disordered system.)

\para
We work here with the formulation of the massive gravity theory \cite{derahm1,derahm2}  presented in \cite{rosen0}. 
 The potential terms for the graviton take a very specific form. They  are constructed using the matrix ${\cal K}$, defined in terms of
the dynamical metric $g_{\mu\nu}$ and a fixed background metric $f_{\mu\nu}$ by
\be {\cal K}^\mu_{\ \rho} {\cal K}^\rho_{\ \nu} = g^{\mu\rho}f_{\rho\nu}\nn\ee
The action for massive gravity is then $S_{\rm bulk}=S_1+S_2$, where $S_2$ describes a two-parameter family of mass-terms
\be
S_2=   \frac{1}{2\kappa^2}\int d^4x\sqrt{-g}\ \Bigg[\alpha  \Tr\,{\cal K}
+ \beta  [(\Tr\,{\cal K})^2- \Tr\,{\cal K}^2]\Bigg]\label{action}\ee
Both $\alpha$ and $\beta$ have dimension of mass${}^2$. As we will see later, perturbations of the 
metric around an ``AdS-like'' background will gain a position dependent mass of the form
\be m^2(r) = -2\beta - \frac{\alpha L}{r}\label{m2}\ee
where $r$ is the radial coordinate, with the UV boundary at $r=0$. For stability of both the bulk and boundary theory, we require $m^2(r)\geq 0$ for all $r$ \cite{vegh,davison}. (In practice this means $m^2(r_h)\geq 0$ where $r_h$ is the IR horizon of the geometry).
%
%

\para
One can construct a massive gravity theory for each choice of background metric $f_{\mu\nu}$. Following \cite{vegh}, we take the degenerate background metric $f_{xx} = f_{yy} =1$ with all other components vanishing\footnote{The proof of the absence of a ghost given in \cite{rosen1,rosen2} assumes that the metric $f_{\mu \nu}$ is invertible. For the degenerate $f_{\mu\nu}$ considered here, it was shown in \cite{vegh} that the theory is ghost free for the $\beta$ mass term but a (seeming) technical limitation means that lack of a ghosts has not yet been proven for  the $\alpha$ mass term. However, as we discuss further in Section~\ref{discuss}, it is likely that ghosts are less of an issue for fiducial metrics that preserve temporal diffeomorphism invariance. }. This means that the action
\eqn{action} continues to enjoy diffeomorphism invariance in the $(r,t)$ coordinates, but there is no diffeomorphism invariance in $(x,y)$ directions. Correspondingly, the boundary theory has a conserved energy, but no conserved momentum currents.

\subsubsection*{Black Brane Backgrounds}

Massive gravity  admits charged black brane solutions  \cite{vegh}. The metric takes the usual form
\be ds^2 = \frac{L^2}{r^2}\left(\frac{dr^2}{f(r)}-f(r)dt^2 + dx^2 + dy^2\right)\label{metric}\ee
For a boundary theory with chemical potential $\mu$, the gauge field is given by 
\be A_t = \mu\left(1-\frac{r}{r_h}\right)\label{a0}\ee
The effect of the graviton mass terms $S_2$ is merely to change the emblackening factor, $f(r)$, a fact which can be simply understood by 
observing that, when evaluated on \eqn{metric}, the mass terms simply play the
role of a position dependent cosmological constant. We have
\be f(r)= 1 + \frac{\alpha L }{2} r + \beta  r^2 - Mr^3 +\frac{\mu^2}{\gamma^2r_h^2} r^4\label{f}\ee
where $\gamma^2 = 2e^2L^2/\kappa^2$.  The requirement of a horizon $f(r_h)=0$ fixes the mass $M$ to be
\be M = \frac{1}{r_h^3}\left(1+\frac{\alpha L r_h}{2} + \beta  r_h^2 + \frac{\mu^2 r_h^2}{\gamma^2}\right)\nn\ee
We will see shortly that, just as in the usual AdS/CFT correspondence, $M$ can be interpreted as the energy density of the
 boundary theory. 

\para
The temperature, $T$, of the boundary theory is easily computed by going to Euclidean signature and requiring that the horizon is smooth with periodic, imaginary time; it is 
\be T = -\frac{f'(r_h)}{4\pi}= \frac{1}{4\pi r_h} \left(3 +\alpha  L  r_h + \beta  r_h^2 - \frac{\mu^2r_h^2}{\gamma^2} \right) \label{t}\ee
%
%
%
%
Further thermodynamic quantities require evaluation of the on-shell bulk action and are sensitive to the counterterms required to remove divergences.  Rather complicated expressions for these quantities were given in \cite{vegh}, but without consideration for the finite contribution of these counterterms. In Appendix A, we show that, with the contribution of these counterterms, the free energy is
\be \Omega \equiv S_{\rm bulk} + S_{\rm boundary} =
-\frac{VL^2}{2\kappa^2}\left(\frac{1}{r_h^3}-\frac{\beta }{r_h} +
\frac{\mu^2}{\gamma^2 r_h}\right) +\epsilon_0(\alpha,\beta)\label{omega}\ee
where $\epsilon_0(\alpha,\beta)$ is an undetermined constant piece which is independent of $T$ and $\mu$ and so  does not affect any thermodynamic quantities.  From a knowledge of $\Omega$, the standard thermodynamics relations follow. The entropy density, $s$, charge density ${\cal Q}$, and energy density ${\cal E}$ are all given by the familiar expressions,
\be
s = \frac{2 \pi}{\kappa^2}\frac{L^2}{r_h^2} \ \ \ \ ,\ \ \ \ {\cal Q} = \frac{\mu}{e^2 r_h} \ \ \ \ ,\ \ \ \ 
{\cal E} = \frac{M L^2}{\kappa^2} + \epsilon_0\label{thermosflask}\ee
The entropy is the usual Bekenstein-Hawking formula, while both charge and energy densities agree with the sub-leading fall-offs of \eqn{a0} and \eqn{f} using the standard AdS/CFT dictionary. This provides evidence that holography using massive gravity is a sensible 
endeavour.  

\para
We note that AdS is not a solution of massive gravity with the degenerate choice of the fiducial metric $f_{\mu\nu}$. Indeed, the ground state, with $T=0$ and $\mu=0$, asymptotes to  AdS${}_2\times {\bf R}^2$ in the infra-red, with the associated finite entropy density (and, in this case, a potentially negative energy density $\sim M$, although this can be compensated by the finite contribution $\epsilon_0$). This means that in the ground state, correlation functions are conformal in the UV while exhibiting local criticality in the IR \cite{semi}.  

\section{Conductivity}

The main purpose of this paper is to explain the universal nature of DC conductivity in holographic massive gravity. A beautifully clear explanation for the universality of transport in holographic models was provided some years ago by Iqbal and Liu \cite{il}. At heart, the idea is that, like happy families, all horizons are the alike. Moreover, certain quantities do not evolve as one moves from the horizon to the boundary of AdS, ensuring that they exhibit universal behaviour. 

\para
We begin this section with a rather extensive review of when universal conductivities arise in general relativity. We then extend these arguments to massive gravity where we will see that they can be applied more broadly.

\subsection*{The Usual Story of Conductivity in GR}

We first consider the usual case of general relativity, with the action given only by $S_{\rm bulk}=S_1$. To compute the conductivity in the regime of linear response, we perturb $\delta A_x$, the spatial component of the gauge field. This then couples to the metric component $\delta g_{tx}$. 
The general expansions of these fields near the boundary are
\begin{eqnarray}
\delta A_x &=& \delta A_x^{(0)} + \frac{r}{L} \delta A_x^{(1)} + ... \nonumber \\
\delta g_{tx} &=& \frac{L^2}{r^2} \delta g_{tx}^{(0)}+ ....\nn
\end{eqnarray}
The leading order term in the gauge field $\delta A_x^{(0)}$ acts as a source for the current. Meanwhile, the leading order term in the metric is a source for the stress tensor and is set to zero: $\delta g_{tx}^{(0)} =0$. We impose ingoing boundary conditions at the horizon, which read
\be \delta A_x \sim
 f(r)^{-i \omega/ 4 \pi T}\label{getin}\ee
At zero momentum, these linear perturbations are governed by the Maxwell equation, 
\be (f(r)\delta A_x')' + \frac{\omega^2}{f(r)}\delta A_x = - \frac{A_t' r^2}{L^2}\left(\delta g_{tx}' +
\frac{2}{r} \delta g_{tx} \right) \nn\ee
together with the  $t-x$ and $r-x$ components of the Einstein equations, which read respectively
\be
\left(\delta g_{tx}' + \frac{2}{r} \delta g_{tx} +\frac{4L^2}{\gamma^2} A_t' \delta A_x\right)' = 0\ \ \ \ {\rm and}\ \ \ \ 
\left(\delta g_{tx}' + \frac{2}{r} \delta g_{tx} + \frac{4L^2}{\gamma^2} A_t' \delta A_x\right) = 0\nn\ee
Clearly the latter  equation implies the former, and we can eliminate $\delta g_{tx}$ to  find a single differential equation for the gauge field. Writing $A_t'=-e^2 {\cal Q}$,  this becomes
\be
(f \delta A_x')' + \frac{\omega^2}{f}\delta A_x = \frac{4 e^4 {\cal Q}^2}{\gamma^2}  r^2 \delta A_x \label{goes}
\ee
Solving this equation, subject to the ingoing boundary conditions \eqn{getin}, allows us to determine the response $\delta A_x^{(1)}$ in terms of the source $\delta A_x^{(0)}$. The ratio is the optical conductivity, which we write as
\be
\sigma(\omega) = \left.\frac{1}{e^2}\frac{\delta A_x'}{i \omega \delta A_x} \right|_{r=0}\label{cond}
\ee
The low frequency behaviour of the conductivity depends crucially on whether there is a background charge density ${\cal Q}$. Mathematically, this difference can be seen in the differential equation \eqn{goes} which, as we now review, has a special  property  when ${\cal Q}=0$. 

\para
The term involving ${\cal Q}$ on the right-hand side of \eqn{goes} acts like a mass term for the gauge perturbation. When ${\cal Q}=0$, the perturbation $\delta A_x$ is effectively massless and,  in the limit $\omega\rightarrow 0$,  equation \eqn{goes} ensures that there is a conserved quantity, $\Pi=f(r)\delta A'_x$, which does not evolve radially: $\Pi'=0$. This is the radial momentum conjugate to $\delta A_x$.
Motivated by this, we define a fictitious membrane DC conductivity living at each constant $r$ slice,
\be
\sigma_{DC}(r) = \lim_{\omega \rightarrow 0}\ \left.\frac{1}{e^2} \frac{\Pi}{i \omega
\delta A_x} \right|_{r}
\ee
At the UV boundary, $r=0$,  this coincides with the $\omega\rightarrow 0$ limit of the boundary theory conductivity \eqn{cond}. Meanwhile, at $r=r_h$, this can be thought of as the DC conductivity of the horizon. The argument of Iqbal and Liu \cite{il} is that $\sigma_{DC}$ does not vary with $r$, ensuring that the conductivity of the boundary theory  is insensitive to details of the bulk geometry and instead  depends only on geometric properties of the horizon, or ``membrane fluid". This follows from the observation that  $\Pi'\sim {\cal O}(\omega^2)$, while  $\delta A_x'\sim \Pi\sim {\cal O}(\omega)$. In particular, at the horizon the ingoing boundary condition \eqn{getin} means that
\be
\Pi = f(r) \delta A_x' = i \omega \delta A_x \sim \mathcal{O}(\omega)
\ee
The net result is that, in the absence of a chemical potential, the DC conductivity can be computed as $\sigma_{DC}=\sigma_{DC}(r_h)$ and is given by,
\be
\sigma_{DC} = \frac{1}{e^2} \frac{\Pi(r_h)}{i \omega \delta A_x(r_h)} = \frac{1}{e^2}\ \ \ \ \ \ \ \ \ \ \ \ (\mathcal{Q}=0)\nn
\ee
This result was previously derived in \cite{cond1}. The formulation above makes it explicit that the conductivity is controlled by the horizon.

\para
In the presence of a chemical potential, the story above no longer holds. Now the  gauge perturbation $\delta A_x$ has an effective mass and there is a non-trivial flow of $\sigma_{DC}$ from the horizon to the boundary. 
%
%
On dimensional grounds one expects $\Pi \sim \mathcal{Q}^2 r_h^3 \delta A_x$ near the boundary. Since $\Pi$ is now $\mathcal{O}(1)$, there is a pole in the imaginary part of the DC conductivity and, by Kramers-Kronig, an accompanying delta-function in the real part,
\be
\sigma \sim \mathcal{Q}^2 r_h^3 \left(\delta(\omega)-\frac{1}{i \omega}\right)\ \ \ \ \ \  \ \ \ \ \ \ \ (\mathcal{Q} \neq 0)\nn
\ee
As we described in the introduction, this is the expected behaviour for the DC conductivity in a system with momentum conservation and  a finite background charge density.

\subsection*{The Story in Massive Gravity}

We now revisit the story of DC conductivity in the framework of massive gravity. Since momentum is no longer conserved, one expects a finite DC conductivity, even when $\mathcal{Q} \neq 0$ \cite{vegh}. We will show that this conductivity is indeed governed by the horizon of the black brane, generalising the results above to finite charge density. 

\para
In massive gravity, the perturbation of the gauge field $\delta A_x$ once again couples to $\delta g_{tx}$ but, as we will see below, it also sources $\delta g_{rx}$. The Maxwell equation is now given by,
\be
\left(f \delta A_x'\right)' + \frac{\omega^2}{f} \delta A_x = -\frac{A_t' r^2}{L^2}\left(\delta g_{tx}' +
\frac{2}{r} \delta g_{tx} - i\omega \delta g_{rx}\right) \nonumber \ee
Meanwhile, the $t-x$ and $r-x$ components of the Einstein equations are no longer equivalent, and  read
\be
\left(\delta g_{tx}' + \frac{2}{r} \delta g_{tx} - i\omega \delta g_{rx} + \frac{4A_t' L^2}{\gamma^2}\delta A_x\right)' &=& \frac{m^2(r)}{f} \delta g_{tx} \label{tx} \\
\left(\delta g_{tx}' + \frac{2}{r} \delta g_{tx} - i\omega \delta g_{rx}  + \frac{4A_t' L^2}{\gamma^2}\delta A_x\right) &=&
- \frac{i f m^2(r) }{\omega} \delta g_{rx} \label{rx}
\end{eqnarray}
where $m^2(r)$ is the position dependent graviton mass that we introduced earlier,
\be
m^2(r) = - 2 \beta - \frac{\alpha L}{r}\nn\ee
%
%
The fact that \eqn{tx} and \eqn{rx} are no longer equivalent is, ultimately, what requires us to turn on the extra component of the metric $\delta g_{rx}$. Nonetheless, the two equations still imply the constraint 
\be
\frac{i \omega m^2(r)}{f} \delta g_{tx} = \left(m^2(r) f \delta g_{rx}\right)'\nn
\ee
which can once again be used to eliminate $\delta g_{tx}$. This means that  the single evolution equation \eqn{goes} of general relativity is replaced by a pair of coupled, ordinary differential equations for $\delta A_x$ and $\delta \tilde{g}_{rx} =
f(r) \delta g_{rx}$
\begin{eqnarray}
 (f \delta A_x')' +\frac{\omega^2}{f}\delta A_x  =  \frac{4e^4{\cal Q}^2}{\gamma^2}r^2 \delta A_x +  \frac{e^2{\cal Q}r^2}{L^2}\frac{m^2}{i\omega} \delta \tilde{g}_{rx} \label{pert1} \\
 \frac{1}{r^2}\left(\frac{r^2 f}{m^2}\left( m^2 \delta \tilde{g}_{rx}\right)'\right)' +
\frac{\omega^2}{f} \delta \tilde{g}_{rx}  =  \frac{4e^2{\cal Q}L^2}{\gamma^2}i\omega \delta A_x + m^2 \delta \tilde{g}_{rx} \label{pert2}
\end{eqnarray}
As we reviewed above, the key step in the universality argument \cite{il} for the low-energy transport was the identification of  a massless mode which does not evolve between the horizon and the boundary. With this in mind, it is instructive to rewrite the fluctuation equations \eqn{pert1} and \eqn{pert2} in the schematic form
\be
\left(\begin{array}{cc} L_1 & 0 \\ 0 & L_2\end{array}\right)
\left(\begin{array}{c} \delta A_x \\ \delta \tilde{g}_{rx}\end{array}\right) +
\frac{\omega^2}{f}\left(\begin{array}{c} \delta A_x \\
\delta \tilde{g}_{rx}\end{array}\right)= {\cal M}\left(\begin{array}{c} \delta A_x \\
\delta \tilde{g}_{rx}\end{array}\right)\label{matrixpert}\ee
where  $L_1$ and $L_2$ are linear differential operators and ${\cal M}$ is a ``mass matrix",
\be {\cal M} = \left(\begin{array}{cc} 4e^4{\cal Q}^2r^2/\gamma^2  & \ \ \ e^2{\cal Q} r^2 m^2/i\omega L^2  \\ 4e^2{\cal Q} L^2i\omega/\gamma^2  &\ \ \  m^2 \end{array}\right)\nn\ee
%
%
%
%
Mathematically, the key point is that $\det {\cal M}=0$. This means that, even when ${\cal Q}\neq0$, there is a 
particular combination of the fields which does not evolve between the horizon and boundary in the 
limit $\omega\rightarrow 0$. The existence of this massless mode for all ${\cal Q}$ is rather surprising. It would be interesting to gain a better understanding of what this means physically. 

\para
Notice that the fields which diagonalise the mass matrix ${\cal M}$ do not diagonalise the derivative terms. Nonetheless, we will now show that the existence of  a massless mode is sufficient to determine the DC conductivity.
Let's now see how this works in more detail. The massless eigenmode of ${\cal
M}$ is
\be \delta \lambda_1 = \left(1+ \frac{4e^4{\cal Q}^2r^2}{\gamma^2m^2}\right)^{-1}\left[\delta A_x -
\frac{e^2{\cal Q}r^2 }{i \omega L^2} \delta \tilde{g}_{rx}\right]\nn\ee
The other mode, with non-vanishing eigenvalue, is
\be \delta \lambda_2 = \left(1+ \frac{4e^4{\cal Q}^2 r^2}{\gamma^2m^2}\right)^{-1}\left[  \frac{4 e^2{\cal Q} L^2
}{\gamma^2m^2}\delta A_x  + \frac{\delta \tilde{g}_{rx}}{i\omega}\right]\nn\ee
From these expressions, we learn that the conductivity of the boundary theory can be extracted from  the UV behaviour of the massless mode $\delta \lambda_1$ alone, 
\be \left.\sigma(\omega) = \frac{1}{e^2}\frac{\delta \lambda_1'}{i \omega \delta \lambda_1}
\right|_{r=0} \label{sigma}\ee
Because $\delta \lambda_1$ and $\delta \lambda_2$  do not diagonalise the derivative terms, the equations of motion expressed in terms of these fields do not quite decouple.  In particular the equation of motion for the
massless mode $\delta \lambda_1$ reads
\be \left[f \left(1 + \frac{4e^4{\cal Q}^2 r^2}{\gamma^2m^2} \right) \delta \lambda_1' -  \frac{e^2{\cal Q}}{ L^2}\frac{fr^4}{m^2}
 \left(\frac{m^2}{r^2}\right)' \delta \lambda_2 \right]' +
\frac{\omega^2}{f} \left(1 + \frac{4e^4{\cal Q}^2 r^2}{\gamma^2m^2}  \right) \delta \lambda_1 = 0 \nn\ee
%
%
%
%
%
and, in the limit $\omega\rightarrow 0$, we learn that there is once again a quantity $\Pi$ which is conserved under radial flow,
\be \Pi=f \left(1 + \frac{4e^4{\cal Q}^2 r^2}{\gamma^2m^2} \right) \delta \lambda_1' -  \frac{e^2{\cal Q}}{ L^2}\frac{fr^4}{m^2}
 \left(\frac{m^2}{r^2}\right)' \delta \lambda_2
 \label{newpi}\ee
 %
%
This motivates us to define a  DC membrane conductivity associated to each radial slice, $r$, by
\be \sigma_{\rm DC}(r) = \left.\lim_{\omega\rightarrow 0}\
\frac{1}{e^2}\frac{\Pi}{i\omega \delta \lambda_1} \right|_{r}\nn\ee
At the boundary $r=0$, this coincides with the $\omega\rightarrow 0$ limit of the conductivity \eqn{sigma}.

\para
The goal now is to show that $\sigma_{DC}$ does not evolve radially.  As in the case of general relativity, we know that $\Pi'\sim {\cal O}(\omega^2)$.  It remains to show that $\delta \lambda'_1\sim {\cal O}(\omega)$. This follows from the requirement that $\Pi\sim {\cal O}(\omega)$, but only if we can ensure that $\delta \lambda_2\sim {\cal O}(\omega)$ too. 
It is straightforward to derive an equation for $\delta\lambda_2$. Importantly, it does not couple directly to $\delta\lambda_1$, but only $\delta\lambda_1^\prime$. This means that it can be written in the form,
\be
L_3 \delta \lambda_2 + p(r) \delta \lambda_2  + \omega^2 q(r)\delta\lambda_2 \sim  \Pi
\ee
where $L_3$ is another differential operator, and $p(r)$ and $q(r)$ are functions that are independent of $\omega$ whose  forms are  unimportant.  Since we are not explicitly sourcing $\delta \lambda_2$, this equation implies that, in the absence of an instability, we must have $\delta \lambda_2 \sim{\cal O}(\omega)$. Hence, from \eqn{newpi}, we can again deduce that $\delta \lambda_1$ is a constant to leading order.

\para
The upshot of this is that, as in the massless case, $\sigma_{\rm DC}(r)$ does not depend
on the radial position, a statement that now holds true even for $\mathcal{Q} \neq 0$.
All that remains is to compute $\sigma_{\rm DC}(r_h)$ at the horizon. This is
easily achieved. We need only remember that both fields obey ingoing boundary
conditions, $\delta A_x\sim \delta \tilde{g}_{rx}\sim f(r)^{-i\omega/4\pi T}$. With this, we find that the $\delta \lambda_2$ term in \eqn{newpi} vanishes at the horizon, while the $\delta \lambda_1'$ term survives. 

\para
The end result is an expression for the DC conductivity which depends only on $r_h$ and the mass of the graviton at the horizon,
\be \sigma_{\rm DC} = \frac{1}{e^2} \frac{\Pi(r_h)}{i \omega \delta \lambda_1(r_h)}=
\frac{1}{e^2}\left(1+ \frac{4 e^4 {\cal Q}^2}{\gamma^2}\frac{ r_h^2}{ m^2(r_h)}\right)\label{result}\ee
This  should be viewed as a generalisation of the earlier result to finite charge density ${\cal Q}$, with the conductivity again determined entirely by horizon quantities.  When $\alpha=0$ and $T=0$, an analytic expression for the DC conductivity was previously derived using an elaborate matching procedure in \cite{davison}; our result agrees in this limit. 

\para
The general form of \eqn{result} is in agreement with that proposed in \cite{hydro}. When the second term dominates, it is expected to give rise to a standard Drude form, a fact which was proven in a certain parameter range in \cite{davison}. In contrast, when the second term fails to be parametrically larger than the first, we have an incoherent metal.

\para 
All temperature dependence enters into the conductivity through factors of the horizon radius. For black brane solutions in our model, $r_h\rightarrow$ constant as $T\rightarrow 0$, and the contribution to the conductivity from momentum relaxation, $\sigma_{DC}-1/e^2$, tends towards a constant at low temperatures. In the next section we will study a dilaton model where one finds powerlaw behaviour for $\sigma_{DC}-1/e^2$.

\para
The DC conductivity can be related to a scattering time $\tau$ by \cite{hydro,sandiego}
%
\be
\sigma_{\rm DC} = \frac{1}{e^2}+\frac{\mathcal{Q}^2}{{\cal E} + P} \tau
\ee
For us, the energy density ${\cal E}$ is given in \eqn{thermosflask}, while the pressure $P=-\Omega/V$ with $\Omega$, the grand canonical potential, given in \eqn{omega}. We therefore identify
%
\be
\tau^{-1} = \frac{\gamma^2}{4 e^2 r_h^2} \frac{ m^2(r_h)}{{\cal E} + P} =  \frac{s}{4 \pi}
\frac{m^2(r_h)}{{\cal E} + P}\label{scatter}
\ee
In \cite{davison}, Davison studied the hydrodynamics of the black brane solutions in massive gravity. He showed that, in the regime $T^2\gg m^2(r_h)$, momentum dissipation is governed by the simple formula,
\be \partial_i T^{ij} = -\tau^{-1} T^{tj}\nn\ee
The scattering time $\tau$ determined from hydrodynamics is identical to the scattering time \eqn{scatter} extracted from our DC conductivity calculation above.

\section{Generalisation}\label{dilsec}

The formula for the DC conductivity depends only on the graviton mass at the horizon, $m^2(r_h)$. Since the key ingredient in the derivation was the existence of a bulk massless mode --- a linear combination of the graviton and gauge field --- one might expect that the result has more general validity. In this section, we confirm this expectation in a model with dilaton coupling.


\para
We replace the Einstein-Maxwell theory \eqn{action1} of the previous section with the action
\be S_3 = \int d^4x\sqrt{-g}\left[\frac{1}{2\kappa^2}\left(R
+\frac{6}{L^2}\right)-\frac{Z(\phi)}{4} F_{\mu\nu}F^{\mu\nu} - \frac{1}{2}
(\nabla \phi)^2  - V(\phi) \right] \nn\ee
As $\phi$ approaches the UV boundary, we have an effective gauge coupling  $Z(\phi(r=0))= 1/e^2$, but this can now be scale dependent. 
We again supplement this with the graviton mass terms \eqn{action}, so the full action is $S=S_3+S_2$. 


\para
We don't need to determine the specific background solution to compute the conductivity. Instead, we merely assume that it takes the form
\be
ds^2 = \frac{L^2}{r^2}\left( -h(r)e^{- 2 \chi(r)} dt^2 + \frac{dr^2}{h(r)}  + dx^2 + dy^2
\right)\nn\ee
with a horizon at $r=r_h$. We further assume that only the $A_t(r)$ component of the gauge field is non-vanishing. The presence of the dilaton means that the electric field is not a constant in space and Gauss' law ensures that  its  profile is simply determined by
\be
e^{\chi} Z(\phi)  A_t'=-e^2{\cal Q}\nn\ee
where ${\cal Q}$ is identified with the  charge density of the dual theory. 

\para 
We now perturb the background solution by  $\delta A_{x}$, $\delta g_{rx}$ and  $\delta g_{tx}$. Since these perturbations are all odd under parity, they do not source the dilaton. Therefore the only modification to the equations of motion are factors of the background dilaton $Z(\phi)$ that appear in all the terms that descend from the Maxwell action. After eliminating $\delta g_{tx}$, the equations of motion can be written in terms of an emblackening factor $f(r) = h(r) e^{-\chi}$ and  perturbation equations \eqn{pert1} and \eqn{pert2} are generalised to
 \begin{eqnarray}
 \frac{1}{Z(\phi)}(f Z(\phi) \delta A_x')' +\frac{ \omega^2}{f} \delta A_x  =   \frac{e^{-\chi}}{Z(\phi)}\bigg( \frac{4 e^4 \mathcal{Q}^2}{\gamma^2} r^2 \delta A_x + \frac{e^2 \mathcal{Q} r^2}{L^2} \frac{m^2}{i \omega} \delta \tilde{g}_{rx} \bigg) \nn \\
 \frac{1}{r^2}\left(\frac{r^2 f}{m^2}\left( m^2 \delta \tilde{g}_{rx}\right)'\right)' +
\frac{\omega^2}{f} \delta \tilde{g}_{rx}  = e^{-\chi} \bigg( \frac{4 e^2 \mathcal{Q} L^2}{\gamma^2} i \omega \delta A_x + m^2 \delta \tilde{g}_{rx} \bigg) \nn
\end{eqnarray}
The mass matrix ${\cal M}$ associated to these perturbations is given by
\be {\cal M} = \left(\begin{array}{cc} 4 e^4 \mathcal{Q}^2 r^2/ \gamma^2 Z(\phi)  & \ \ \  e^2 \mathcal{Q} r^2 m^2 / i \omega L^2 Z(\phi)  \\   4 e^2 \mathcal{Q} L^2 i \omega/ \gamma^2 &\ \ \  m^2 \end{array}\right)\nn\ee
Importantly, it remains degenerate, with $\det{\cal M}=0$. The argument now proceeds in the same manner as the previous section.  The massless eigenmode of ${\cal M}$ is  given  by
\be \delta \lambda_1 = \left(Z(\phi)+ \frac{4e^4{\cal Q}^2r^2}{\gamma^2m^2}\right)^{-1}\left[Z(\phi) \delta A_x -
\frac{e^2{\cal Q}r^2 }{i \omega L^2} \delta \tilde{g}_{rx}\right]\nn\ee
whilst the massive mode is
\be \delta \lambda_2 = \left(Z(\phi)+ \frac{4e^4{\cal Q}^2 r^2}{ \gamma^2m^2}\right)^{-1}\left[  \frac{4 e^2{\cal Q} L^2
}{\gamma^2m^2}\delta A_x +  \frac{\delta \tilde{g}_{rx}}{i\omega}\right]\nn\ee
The information about the DC conductivity is again contained in the massless mode $\delta \lambda_1$. In the $\omega\rightarrow 0$ limit, it is simple to check that there is  conserved momentum flux
%
%
%
\be
\Pi = f \left( Z(\phi) + \frac{4 e^4 \mathcal{Q}^2 r^2}{ \gamma^2 m^2}\right) \delta \lambda_1' - \frac{e^2 Q}{ L^2} \frac{f r^4}{m^2} \left( \frac{m^2 Z(\phi)}{r^2} \right)' \delta \lambda_2 
\ee
From hereon, the steps are the same as the previous section. We find that the DC conductivity is given by
\be
\sigma_{DC} = \frac{1}{e^2} \left( Z(\phi(r_h)) + \frac{4 e^4 \mathcal{Q}^2 }{\gamma^2} \frac{r_h^2}{m^2(r_h)} \right)
\ee
Only the first term differs from \eqn{result}; this can be simply understand as the renormalization of the vacuum conductivity of the boundary theory. The second term, due to the charge density, remains unaffected by the dilaton. The scattering time $\tau$  extracted from the conductivity is again given by \eqn{scatter}. 

\para
If we work to leading order in $m^2$,  thermodynamic quantities can be computed on the background of a solution in general relativity. One effect of a runaway dilaton is to ensure that the black hole horizon scales with a power of $T$ at low temperatures (see, for example, \cite{blaise}). Correspondingly, the scattering time $\tau$ also exhibits powerlaw dependence, a fact observed in a slightly different context in impurity scattering  \cite{pauli}.  After this paper appeared, this system was subsequently explored in more detail in \cite{richard2}.

%
%
%
 

\section{Discussion}
\label{discuss}

\enlargethispage{5pt}
The main result of this paper is a simple analytic formula for the conductivity \eqn{result} in massive gravity. This result is universal in the sense that it depends only on the value of the graviton mass at the  horizon. Moreover, as we have seen in Section \ref{dilsec}, it holds in a large class of theories and is essentially unchanged by the presence of a dilaton field. 

\para
However, we should emphasise that this result is less robust than the famous $\eta/s=1/4\pi$ calculation. In the case of shear viscosity, the universality relied not only on the absence of radial flow, but also on the universality of the coupling to a transverse graviton. In contrast, our result for the DC conductivity can be affected by changes to both the electromagnetic sector and the gravitational sectors of the bulk theory. We now discuss each of these in turn.

\para
In the electromagnetic sector, we restricted attention to theories in which the entire charge density is hidden behind the horizon.   If we were
to instead include charged matter in the bulk, then an explicit current would be generated by the electric field. The structure of the Maxwell equations would
be modified by the $J^x A_{x}$ coupling in the action and it is not clear how to generalise our argument.

\para Of course a breakdown in our argument in the presence of explicitly charged matter is required to explain the onset of a superconducting phase transition. It is simple to understand how this arises within massive gravity. The presence of a charged condensate will give an extra
mass term to the fluctuation $\delta A_x$ and  the mass matrix will no longer have vanishing determinant. This is sufficient to restore some component of the delta-function in the zero frequency conductivity. 

\enlargethispage{5pt}

\para
It is perhaps more interesting to ask how changes to the gravitational dynamics affect the result. 
In this paper, we have worked with the form of  $m^2(r)$ given in \eqn{m2} that was argued in \cite{derahm1,derahm2,rosen1,rosen2} to be the unique ghost-free version of massive gravity. However, in the context of holography there appears to be a loophole. The arguments in the papers above all apply to a non-degenerate fiducial metric $f_{\mu\nu}$. But, as we've seen, to implement holographic momentum dissipation we need to work with the degenerate fiducial metric with vanishing timelike components. Since ghosts typically arise from timelike components of gauge fields and metrics, it seems plausible that the surviving temporal diffeomorphism invariance will allow  more general graviton potential terms of the form $\Tr\,{\cal K}^n$ for $n\geq 3$, without the rise  of re-introducing the Boulware-Deser ghost\footnote{An alternative way to change the effective graviton mass $m^2(r)$ is to change the background fiducial metric to a function of position: $f_{xx}=f_{yy}=F(r)$. This has the consequence of removing diffeomorphism invariance in the radial direction, which may make one nervous about the holographic nature of the bulk. (See however \cite{don} for a recent discussion of what is necessary to make holography work). }. We stress that the form of the DC conductivity \eqn{result} would continue to hold even in this more general construction of massive gravity.


 \para
 Perhaps the most pressing issue is to try to derive some form of massive gravity through  an exact treatment of a system which exhibits momentum dissipation, whether through impurities, a background lattice, or coupling to a larger bath of particles. 
  The background solution is presumably determined by the original, complicated theory and need not be related to massive gravity. 
 However, it may be that the perturbations around this background coincide with those of massive gravity, now with a more general function $m^2(r)$ for the graviton mass. Since the conductivity formula \eqn{result} cares nothing for the background geometry, it still holds in this more general context, with the DC conductivity determined only by the graviton mass at the horizon. If true, the formula \eqn{result} would have a much larger realm of applicability than the specific massive gravity theory discussed in this paper\footnote{These expectations were borne out in \cite{lattices}.}.

\para Needless to say, it would be interesting to place the above speculations on a firmer footing and determine the role played by massive gravity in more microscopic examples of holography with momentum dissipation.

\pagebreak

 %
%



\appendix

\section{Appendix: Thermodynamics}

Rather complicated expressions for the thermodynamics of black branes in massive gravity were presented in \cite{vegh}. These were derived by computing the on-shell action for the Euclidean black brane and subtracting the divergent terms. However, even in massless gravity, a proper treatment of the gravitational counterterms gives rise to extra finite contributions to the free energy which can modify the thermodynamics. For general relativity, the bulk Euclidean action is augmented by the counterterm
\be
S_{{\rm counter}} = \frac{1}{2\kappa^2}\int \mathrm{d}^3x \sqrt{\gamma} \left(- 2\Theta + \frac{4}{L}\right)
\label{grcounter}\ee
where $\gamma_{\mu \nu}$ is the induced metric on the boundary and $\Theta =
\gamma^{\mu \nu} \nabla_{\mu} n_{\nu}$ ,with $n_{\nu}$ an outward facing normal
to the boundary, is the extrinsic curvature. The first term is nothing but the
usual Gibbons-Hawking-York boundary term, required to have a well-defined
variational principle in the bulk. The second term removes divergences
associated with asymptotically AdS spaces.  For charged black holes in Einstein-Maxwell theory, the net effect of the counterterm is merely to change thermodynamic expressions by a factor of 2.

\para 
In the case of massive gravity, the space-time is not asymptotically locally AdS in the usual sense; there are now linear and quadratic terms
in $f(r)$ for all solutions. This has the consequence that the counterterm \eqn{grcounter} is insufficient in massive gravity: one needs to add new counterterms proportional to the graviton mass parameters $\alpha$ and $\beta$. A correct derivation of these counterterms would require that the on-shell action is finite when evaluated on all solutions. We do not do this analysis here; it will suffice for our purposes to simply deduce the counterterms necessary to remove the divergences of the black hole solution \eqn{metric}. As we will see, up to a reasonable assumption (to be specified below) this will allow us to determine the thermodynamics.

\para
In a general analysis, the counterterms should be written in terms of boundary fields. However, when evaluated on the solution \eqn{metric}, the form of the counterterms at $r=\epsilon$ must be given by
\be S'_{{\rm counter}} = \frac{1}{2\kappa^2}\int \mathrm{d}^3x \sqrt{\gamma} \left(- 2\Theta + \frac{4}{L}\left[A+B\frac{\epsilon}{L}+C\frac{\epsilon^2}{L^2}+ D\frac{\epsilon^3}{L^3}+\ldots\right]\right)\label{mcounter}\nn\ee
where $\ldots$ are terms which vanish as $\epsilon\rightarrow 0$.

\para
We require that $S=S_{\rm bulk}+S'_{\rm counter}$ is finite when evaluated on \eqn{metric}. (Here $S_{\rm bulk}$ should be the Euclidean bulk action, which differs by an overall minus sign from \eqn{action1}). This is sufficient to uniquely determine the coefficients $A$, $B$ and $C$;
\be
A=1\ \ \ ,\ \ \ B= \frac{\alpha L^2}{4} \ \ \ ,\ \ \ C = \frac{\beta L^2}{2} - \frac{\alpha^2L^4}{32}\nn\ee
Note that neither $B$ nor $C$ depend on $T$ and $\mu$. This is unsurprising since their role is to cancel the divergences arising from the linear and quadratic terms in $f(r)$ that are due to massive gravity. Moreover, in contrast to the $\Theta$ and $A$ terms, they do not affect the thermodynamics, contributing only a ($\alpha$ and $\beta$ dependent) constant to the free energy.

\para
The requirement of a finite on-shell action does not fix the coefficient $D$. Here we simply assume that, like $B$ and $C$, this too depends only on $\alpha $ and $\beta$ and has no explicit dependence on $T$ or $\mu$. With this assumption, its role is only to add a constant $\epsilon_0(\alpha,\beta)$ to the free energy\footnote{We note that the form of $B$ and $C$ is suggestive that the full counterterm, evaluated on the solution, takes the form
\be S'_{\rm counter}  = \frac{1}{2\kappa^2}\int \mathrm{d}^3x \sqrt{\gamma} \left(- 2\Theta + 4\sqrt{\frac{1}{L^2} + \frac{\alpha\epsilon}{2L} + \frac{\beta \epsilon^2}{L^2}}\right)\nonumber\ee
Such square-root forms of counterterms, typically involving the boundary Ricci scalar,  have been suggested previously \cite{lau,mann}.  Taking this as a hint for the value of $D$ gives $\epsilon_0(\alpha,\beta)=0$ in \eqn{theanswer}.}.

\para
With these counterterms in place, the on-shell bulk action is, by construction, finite and given by
\be \Omega \equiv S_{\rm bulk} + S'_{\rm counter} =
-\frac{VL^2}{2\kappa^2}\left(\frac{1}{r_h^3}-\frac{\beta}{r_h} +
\frac{\mu^2}{\gamma^2 r_h}\right)+\epsilon_0(\alpha,\beta)\label{theanswer}\ee
where $V$ denotes the (infinite) spatial area of the boundary theory. As usual
in AdS/CFT, $\Omega$ is identified with the potential in the grand canonical
ensemble. From this we can calculate the entropy density,
\be s = -\frac{1}{V}\frac{\partial \Omega}{\partial T} =
\frac{2\pi}{\kappa^2}\frac{L^2}{r_h^2}\nn\ee
Happily this is precisely the Bekenstein-Hawking entropy. The charge density is
given by
\be {\cal Q} = -\frac{1}{V}\frac{\partial \Omega}{\partial \mu} = \frac{\mu}{e^2
r_h}\nn\ee
as expected. Finally, the energy density is equal to what we would usually call
the mass of the black brane, supplemented by the additional, undetermined $\epsilon_0(\alpha,\beta)$,
\be {\cal E} = \frac{\Omega}{V} + sT + {\cal Q} \mu = \frac{M L^2}{\kappa^2} + \epsilon_0\nn\ee

\section*{Acknowledgements}

We would like to thank Allan Adams,  Aristos Donos, Andreas Karch, Elias Kiritsis, Don Marolf, Kostas Skenderis, Henry Tye, Jan Zaanen and especially Richard Davison, Sean Hartnoll and David Vegh for useful discussions and comments.   We are supported by STFC and by the European
Research Council under the European Union's Seventh Framework Programme
(FP7/2007-2013), ERC Grant agreement STG 279943, Strongly Coupled Systems


\end{document}